\documentclass[twocolumn,amsmath,amssymb,prb]{revtex4}

\usepackage{graphicx}

\usepackage[breaklinks]{hyperref}
\hypersetup{
    bookmarks=false,
    pdfstartview={FitH},
    colorlinks=true,
    linkcolor=blue,
    citecolor=blue,
    urlcolor=blue
}

\begin{document}

\title{Electronic structure of the kagom\'{e} staircase compounds
Ni$_3$V$_2$O$_8$ and Co$_3$V$_2$O$_8$}

\author{J.~Laverock}
\author{B.~Chen}
\author{A.~R.~H.~Preston}
\author{K.~E.~Smith}
\affiliation{Department of Physics, Boston University, 590 Commonwealth Avenue,
Boston, MA 02215, USA}

\author{N.~R.~Wilson}
\author{G.~Balakrishnan}
\affiliation{Department of Physics, University of Warwick, Coventry, CV4 7AL,
United Kingdom}

\author{P.-A.~Glans}
\author{J.-H.~Guo}
\affiliation{Advanced Light Source, Lawrence Berkeley National Laboratory,
Berkeley, CA 94720, USA}

\begin{abstract}
The electronic structure of the kagom\'{e} staircase compounds,
Ni$_3$V$_2$O$_8$ and Co$_3$V$_2$O$_8$, has been investigated using soft x-ray
absorption, soft x-ray emission, and resonant inelastic x-ray scattering
(RIXS). Comparison between the two compounds, and with first principles band
structure calculations and crystal-field multiplet models, provide unquie
insight into the electronic structure of the two materials. Whereas the location
of the narrow (Ni,Co) $d$ bands is predicted to be close to $E_{\rm F}$,
we experimentally find they lie deeper in the occupied O $2p$ and unoccupied
V $3d$
manifolds, and determine their energy via measured charge-transfer excitations.
Additionally,
we find evidence for a $dd$ excitation at 1.5~eV in Ni$_3$V$_2$O$_8$,
suggesting the V $d$ states may be weakly occupied in this compound, contrary
to Co$_3$V$_2$O$_8$. Good agreement is found between the crystal-field $dd$
excitations observed in the experiment and predicted by atomic multiplet
theory.
\end{abstract}

\maketitle

\section{Introduction}
The coupling between spin, charge and lattice degrees of freedom in metal
oxides yields rich phase diagrams of competing phases, and has long attracted
significant attention. The kagom\'{e} staircase family of vanadium oxides, {\em
M}$_3$V$_2$O$_8$ ({\em M}~=~transition metal), support complex low-temperature
magnetic phase diagrams including incommensurate spin structure,
and multiferroic behavior.
\cite{lawes2005,wilson2007etc}
While there have been numerous studies of
the cascade of low-temperature magnetic phases in these oxides, there
have been relatively few investigations of their fundamental
electronic structure.
Recently, detailed field- and temperature-dependent optical studies
on Ni$_3$V$_2$O$_8$ (NVO) and
Co$_3$V$_2$O$_8$ (CVO) were combined with {\em ab initio} band structure
calculations, and revealed the inadequacy of the
local (spin) density approximation [L(S)DA] in describing the electronic
structure of these materials. \cite{rai2006,rai2007}
Moreover, inclusion of electron correlations,
in the form of the LDA+U method, were not found to significantly improve
agreement with experiment.

We report here a comprehensive soft x-ray spectroscopic study of the electronic
structure of Ni$_3$V$_2$O$_8$ and Co$_3$V$_2$O$_8$ where we
directly measure both the occupied and excited states
through absorption, emission and resonant inelastic
x-ray scattering measurements. Comparisons are made with {\em ab initio}
band theory and crystal-field multiplet calculations, as well as between the
two compounds, to provide a detailed description of their electronic structure.

\begin{figure}[b!]
\begin{center}
\includegraphics[width=0.8\linewidth,clip]{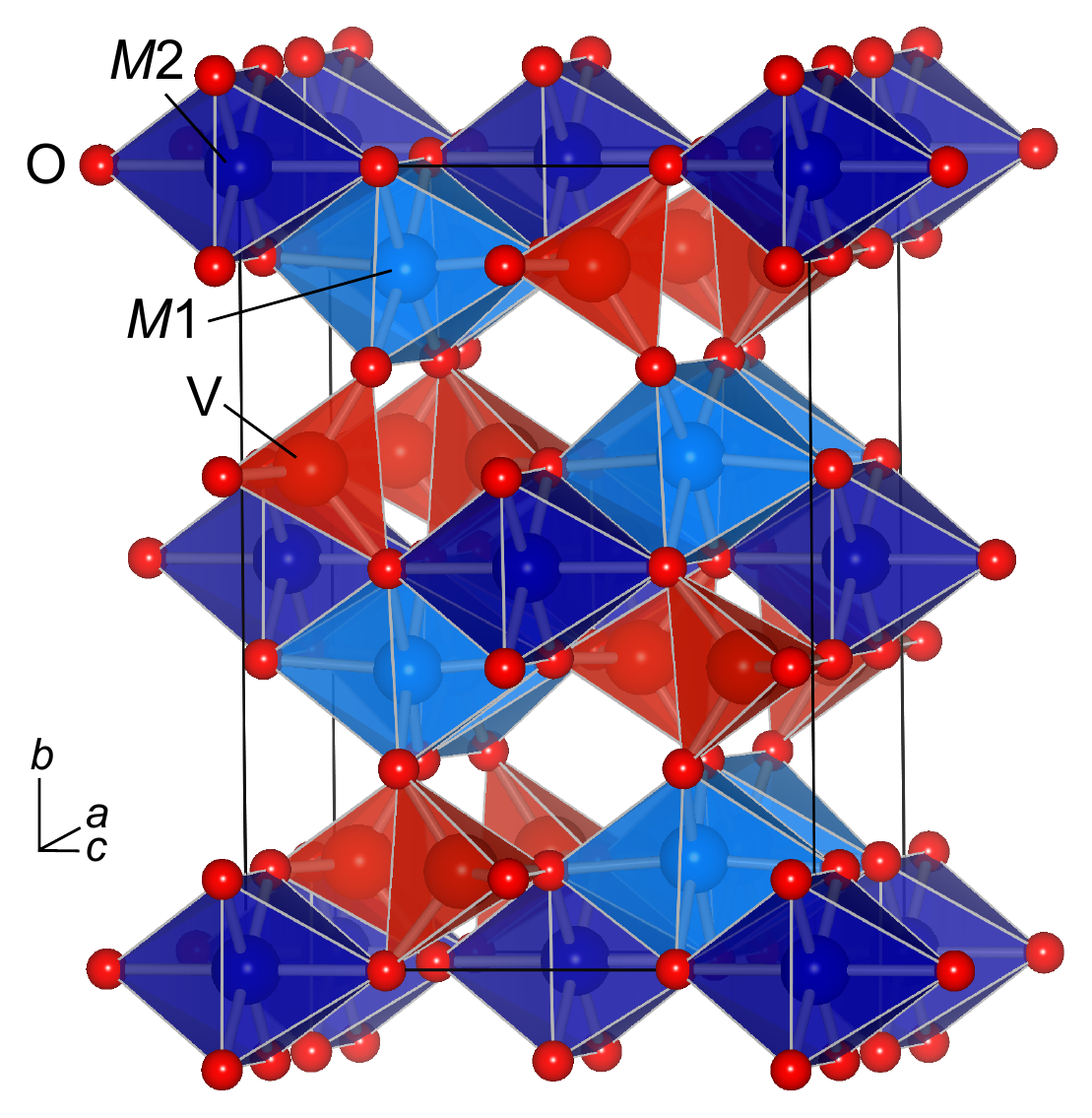}
\end{center}
\vspace*{-0.2in}
\caption{\label{f:struc} (Color online) Crystal structure of {\em
M}$_3$V$_2$O$_8$, consisting of {\em M}O$_6$ octahedra and VO$_4$
tetraheda. The two inequivalent {\em M} sites are displayed in light ({\em
M}1, cross-tie) and dark ({\em M}2, spine) blue.}
\end{figure}

{\em M}$_3$V$_2$O$_8$ ({\em M}~=~Co, Ni) crystallizes in the centered
orthorhombic space group {\em Cmca}, consisting of
edge sharing {\em M}$^{2+}$O$_6$ octahedra and V$^{5+}$O$_4$ tetrahedra
(see Fig.~\ref{f:struc}). \cite{sauerbrei1973}
Two inequivalent {\em M} sites exist within
this structure, which are commonly referred to as cross-tie ({\em M}1, two
sites per unit cell) and spine ({\em M}2, four sites per unit cell) sites.
Below $\approx 10$~K, a series of ordered magnetic phases develop, which
vary depending on the {\em M} ion. Generally, high-temperature
incommensurate magnetic structures eventually give way to a commensurate ground
state. For NVO, the broken inversion symmetry of one of the incommensurate
spin structures brings the coupling of the magnetism to ferroelectricity
into play, yielding coupled multiferroic behavior. \cite{lawes2005}

Soft x-ray absorption spectroscopy (XAS) and soft x-ray emission spectroscopy
(XES) are element-specific probes of the electronic structure of complex
materials. In the XAS process, an electron is excited from a core level to
the unoccupied conduction band states, subject to dipole selection rules
($\Delta{l} = \pm 1$). \cite{stohr1992}
In the case of XAS at the O $K$-edge, this involves
transitions between O $1s$ to O $2p$ states. The hybridization between O $2p$
wavefunctions and neighboring transition metal $d$ electrons makes O $K$-edge
XAS ideal as a probe of the (hybridized) unoccupied $d$ states in transition
metal oxides, allowing direct comparison with the partial density of states
(PDOS)
obtained from first-principles band theory. \cite{stohr1992}
At the transition metal $L$-edge
({\em M} $2p \rightarrow 3d$), the large overlap between the $2p$ core hole
and $3d$ wavefunctions mean that atomic multiplet effects dominate these
spectra, overwhelming band effects. In normal fluorescent XES, the system is
initially excited in a process similar to the XAS process, but the measurement
focuses on the fluorescent decay of the system (that fills the core hole).
\cite{smith2003} For
example, at the O $K$-edge, one measures $2p \rightarrow 1s$ transitions,
corresponding directly to the occupied O PDOS. However, when the initial
excitation is tuned to an edge
feature of the absorption spectrum, resonant effects become important,
and the incident photon transfers energy and momentum to the system.  Such a
measurement is referred to as resonant inelastic x-ray scattering (RIXS), and
the emitted photon carries information about the low-energy excitations.
\cite{ament2011b} As
well as probing delocalized excitations (such as magnons or orbitons), RIXS
can yield information on local crystal-field transitions ($dd$ transitions)
and charge-transfer (CT) transitions.

\section{Methods}

\subsection{Experimental}
High-quality single crystals of Ni$_3$V$_2$O$_8$ and Co$_3$V$_2$O$_8$
were grown using the floating zone method with an optical image furnace.
\cite{balakrishnan2004} Samples were cleaved
{\em ex-situ}, immediately before loading into the ultra-high vacuum chamber.
Soft x-ray spectroscopy measurements were performed at room temperature
(i.e.~in the paramagnetic insulating phase) at
Beamline 7.0.1 at the Advanced Light Source (ALS), Lawrence Berkeley National
Laboratory and at Beamline X1B
of the National Synchrotron Light Source, Brookhaven National Laboratory.
XAS measurements were
made in both total electron yield (TEY) and total fluorescent yield (TFY)
modes. At the V $L$- and O $K$-edges, the energy resolution was set to 0.2~eV
at full width at half maximum (FWHM), and the photon energy was calibrated
with reference to TiO$_2$ Ti $L$- and O $K$-edge spectra. At the Co and Ni
$L$-edges, the energy resolution was set to 0.2~eV and 0.3~eV respectively,
and the photon energy was calibrated with CoO and NiO $L$-edge spectra. Of
the two XAS modes, TEY is more surface sensitive, having a probing depth of
$\sim 10$~nm compared with the $\sim 100$~nm probing depth of TFY. In all
the measurements, good correspondence was found between TEY and TFY spectra,
indicating our results are representative of the bulk electronic structure. XES
spectra were recorded with a Nordgren-type spectrometer, \cite{nordgren1989}
with energy resolution between 0.7 and 1.0~eV (depending on the emission
feature under
study), and the instrument was calibrated with reference to metallic Zn,
Co and Ni $L_{3,2}$-edge spectra. All emission measurements were performed
with a $90^{\circ}$ angle between incident and scattered photons, with the
polarization vector parallel to the horizontal scattering plane, and with an
angle of $\approx 70^{\circ}$ between the incident photons and the surface
normal of the sample (i.e.~near-grazing geometry).

\subsection{Band structure calculations}
For comparison with our experimental results, {\em ab initio} band structure
calculations have been performed using the all electron full-potential
linearized augmented plane-wave (FLAPW) method, as implemented in the
{\sc Elk} code. \cite{elk} The deficiencies of the LSDA in describing the
electronic structure of {\em M}$_3$V$_2$O$_8$ have been well established
through detailed FLAPW calculations; for example, CVO
is predicted to be metallic. \cite{rai2006,rai2007} However, those authors
also found that although
the inclusion of static Coulomb correlations (in the form of the Hubbard
$U$ parameter) produced an insulating ground state for CVO, the magnitude
of the band gap was inconsistent with experiment for both NVO and CVO, and
the predicted optical spectra were found to be in no better agreement with
experiment. Here, we employ model calculations within the LSDA of both CVO and
NVO, focusing on the bonding and hybridization characteristics, to study the
{\em trends} between the two materials. The crystallographic parameters used
in the calculations are those reported in the literature. \cite{sauerbrei1973}
Self-consistency in the calculation was achieved on 125 k-points in the
irreducible ($1/8^{\rm th}$) Brillouin zone, with a cutoff for plane waves
in the interstitial region of $k_{\rm max} = 7.5/R_{\rm min}$, where $R_{\rm
min}$ is the O muffin-tin radius. Muffin-tin radii of 2.2~a.u., 1.55~a.u. and
1.55~a.u. were used for {\em M}, V and O respectively.

\subsection{Crystal-field multiplet calculations}
RIXS at the transition-metal $L$-edge is dominated by local crystal-field
transitions between $3d$ configurations. \cite{kotani2001,ament2011b}
In order to complement our RIXS measurements, crystal-field multiplet (CFM)
calculations have been performed using the {\tt CTM4XAS} and {\tt CTM4RIXS}
programs. \cite{stavitski2010}  In contrast to other schemes, such as the
single-impurity Anderson model (SIAM), the CFM approach takes no direct account
of the ligand. For this reason, the value of the crystal-field parameter,
$10Dq$, used in CFM calculations represents the effective separation between
$t_{2g}$ and $e_g$ states, which includes (approximately and empirically)
the effects of hybridization with the ligand. However, a direct comparison
between SIAM and CFM calculations at the Mn $L$-edge of MnO has demonstrated
the good correspondence between both theories, with a renormalization
of the CFM $10Dq$ value by a factor of two compared with the SIAM value.
\cite{ghiringhelli2006} More obviously, a second consequence of omitting
the effects of the ligand is that charge transfer excitations cannot be
accounted for in the CFM model.

\subsection{Data analysis}
The maximum entropy (MaxEnt) approach has recently been shown to be a
valuable tool in deconvoluting x-ray emission spectra, \cite{laverock2011b}
and is employed here in the Co and Ni $L$-edge RIXS presented in
Sections~\ref{s:cvo_col_rixs} and \ref{s:nvo_nil_rixs}. However, owing to the
favorable propagation of noise through the MaxEnt procedure,
\cite{laverock2011b} the MaxEnt process can also be
used as an efficient noise filter. First, the raw spectrum is deconvoluted as
usual, using the appropriate resolution function. Second, the deconvoluted
spectrum is convolved with the same resolution function. The result of this
process is a spectrum with the same resolution as the original measurement,
but with an improved signal-to-noise ratio.  This approach
is used in the V $L$-edge RIXS measurements of Section~\ref{s:cnvo_vl_rixs},
owing to the weak scattering intensity of these $d^0$ compounds at this edge.

In the RIXS measurements, the scattered photons are measured using a
two-dimensional (2D) detector, in which each horizontal slice on the detector
represents an equivalent intensity-energy spectrum. \cite{nordgren1989}
Owing to the curvature of
the 2D detector inherent in the measurement, the energy scales are slightly
different, and higher fidelity spectra can be obtained by sampling the data
at sub-pixel channel widths. This approach, made possible by the curvature,
has been employed in all of the RIXS measurements
that follow.

\begin{figure}[t!]
\begin{center}
\includegraphics[width=0.8\linewidth,clip]{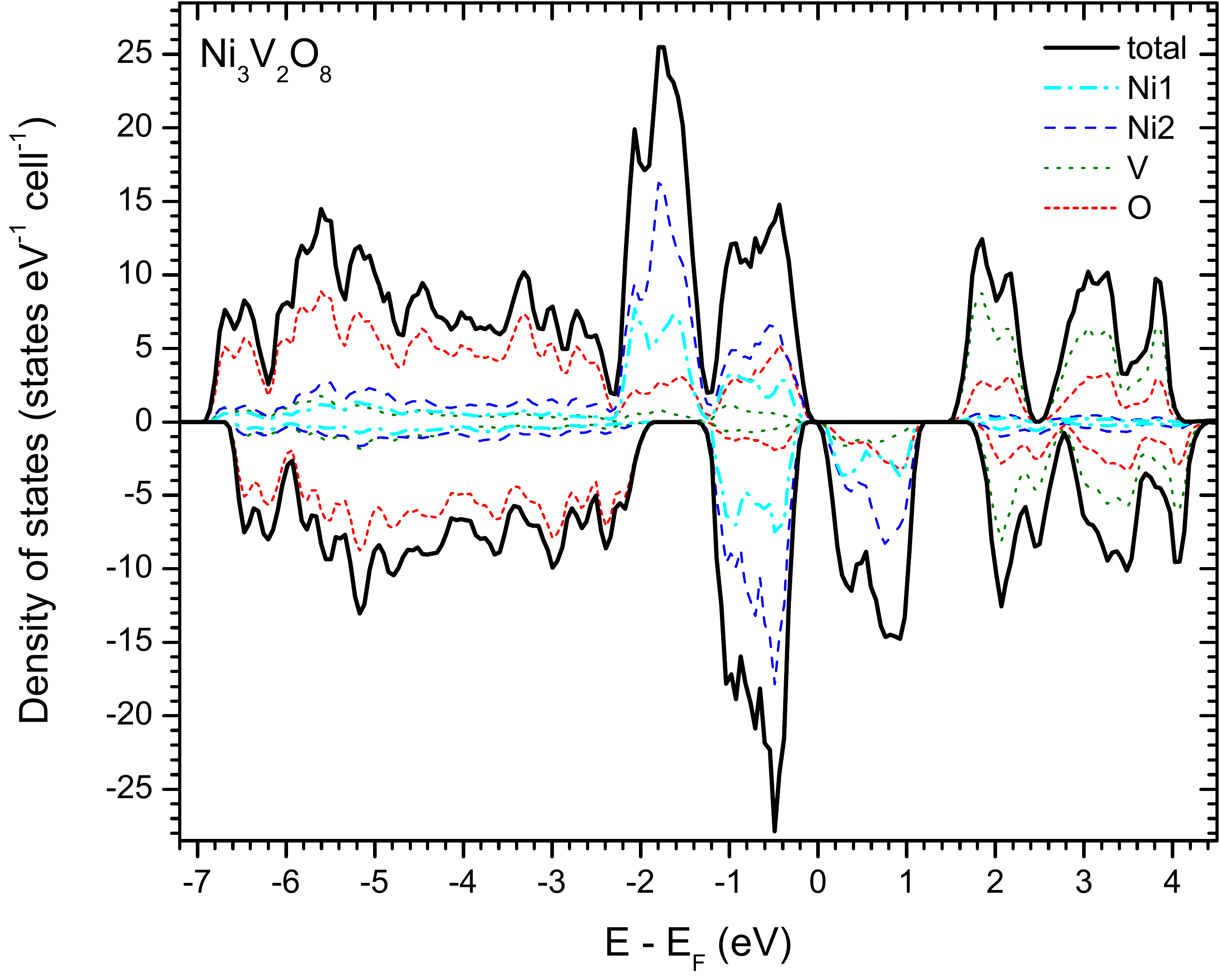}
\includegraphics[width=0.8\linewidth,clip]{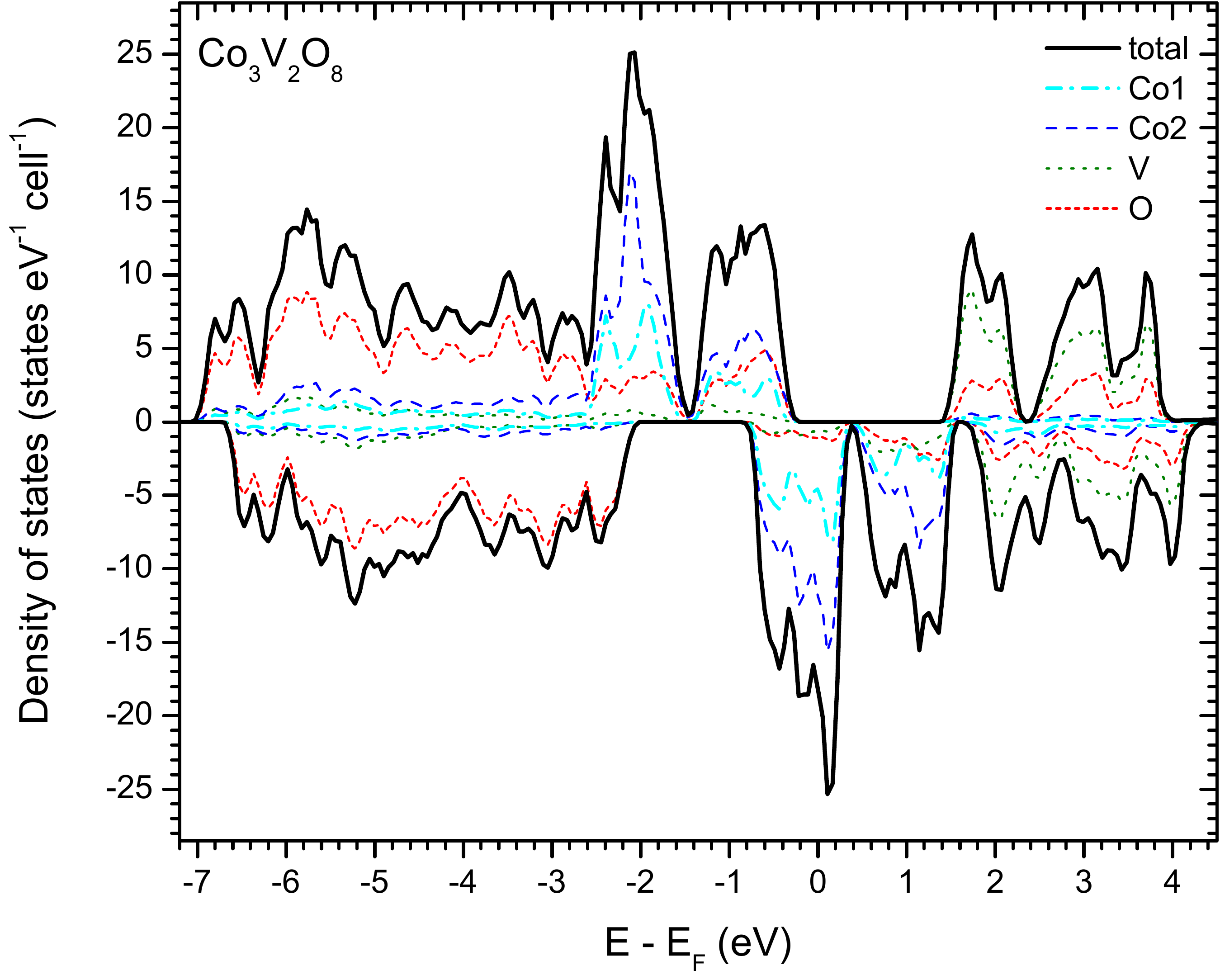}
\end{center}
\vspace*{-0.2in}
\caption{\label{f:cnvo_elk} (Color online) Partial densities of states of
ferromagnetic NVO and CVO from FLAPW calculations.}
\end{figure}

\section{Results}
\subsection{Band structure calculations}
Since both NVO and CVO are local moment insulators, we compare our experimental
results with ferromagnetic electronic structure calculations, since such
calculations correctly predict the occupation of the
(Ni and Co) $3d$ states, and the insulating nature of NVO. This is in contrast
to paramagnetic (degenerate spin) calculations.
In fact, through comparison with the corresponding
paramagnetic solutions, we find that the V and O states (including their
energies and hybridization) are relatively insensitive to the magnetism
for our purposes. The net effect of different magnetic orderings, such as
ferrimagnetic, have been shown to have only weak effects on the overall density
of states. \cite{rai2006,rai2007} As expected, our ferromagnetic calculations
yield magnetic moments of $\sim 2$~$\mu$B per Ni and $\sim 3$~$\mu$B per Co,
corresponding to the high-spin $3d$ occupations of $t_{2g}^6\;e_{g\uparrow}^2$
and $t_{2g}^5\;e_{g\uparrow}^2$ respectively, and are in good agreement with
the LSDA calculations of Rai~{\em et al.} \cite{rai2006,rai2007} The
calculated densities
of states of the two compounds are shown in Fig.~\ref{f:cnvo_elk}. For NVO,
a small band gap between $e_{g\uparrow}$ and $e_{g\downarrow}$ indicates the
insulating ground state, whereas $E_{\rm F}$ is incorrectly placed in the
middle of the $e_{g\downarrow}$ manifold for CVO. With the inclusion of static
electron correlations, in the form of the LSDA+U method (not investigated
here), the occupied (Ni,Co) bands were found to be forced into the O $2p$
manifold, whereas the empty (Ni,Co) states were pushed upwards into the V $3d$
states. \cite{rai2006,rai2007} Whilst this correctly reproduces the insulating
character of the ground state, it was pointed out by those authors that the
magnitude of the band gap was substantially overestimated for reasonable
values of $U$, and that the agreement with optical spectra was worse.

\begin{figure}[t!]
\begin{center}
\includegraphics[width=1.0\linewidth,clip]{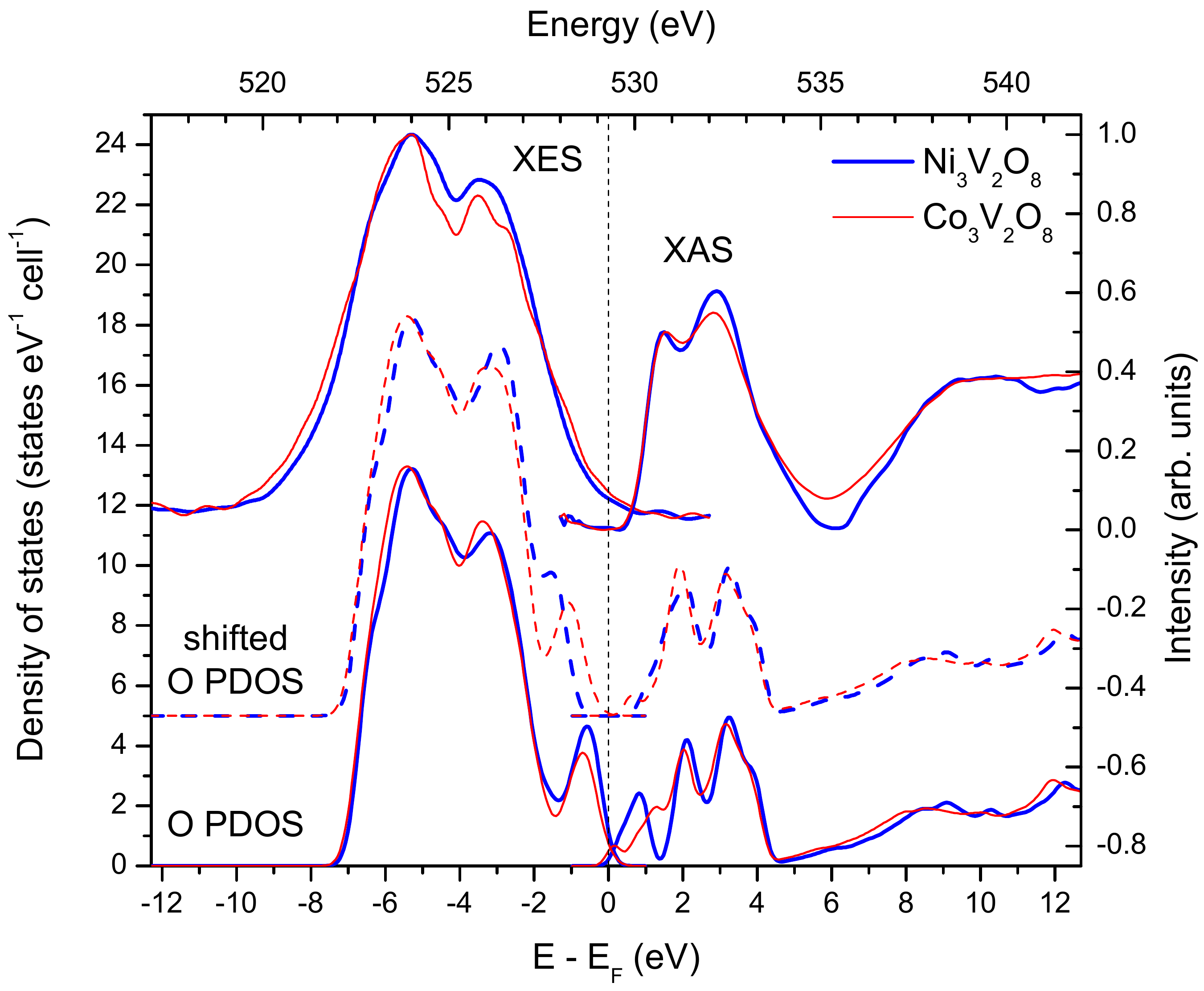}
\end{center}
\vspace*{-0.2in}
\caption{\label{f:cnvo_pdos} (Color online) O $K$-edge XES and XAS
measurements of CVO and NVO compared with the spin-averaged O PDOS from the
FLAPW calculations. The TFY-XAS spectra have been shifted to higher energies by
1~eV to account for the effect of the core-hole.  The PDOS has been broadened
by 0.6~eV (occupied) and 0.4~eV (unoccupied) for comparison with the data,
and is shown rigidly shifted to match with the O $K$-edge XES spectra. The
dashed lines represent the PDOS after artificially shifting the $d$ states
away from $E_{\rm F}$ (see text).}
\end{figure}

\subsection{O $K$-edge XAS/XES}
In Fig.~\ref{f:cnvo_pdos}, we compare our experimental XES and XAS results
with the calculated band structure. As previously discussed, XES and XAS
at the O $K$-edge reflect the occupied and unoccupied
O $2p$ PDOS, respectively.
Note that the XAS spectrum has been shifted to higher
energies by 1~eV to (approximately) account for the effects of the core hole.
\cite{preston2008}  After this correction,
the spectra are consistent with a small gap between hybridized states
of 0.5~--~1.0~eV in both compounds, which is in rough agreement
with the optical gap of $\sim 0.4$~eV. \cite{rai2006,rai2007} Note that pure $d$
states that do not hybridize with O will not be visible in these spectra.
In general,
the two compounds show remarkable similarity in their experimental spectral
features, although CVO is found to have a slightly broader O $2p$ bandwidth,
with a tail that extends closer to $E_{\rm F}$. Overall, reasonably good
agreement is observed between experiment and theory in the O $2p$ ($-7$
to $-2$~eV) and V $3d$ (2 to 4~eV) manifolds. In the O $2p$ manifold,
two features are well resolved in the XES experiment: the lower peak at
524~eV originates from the oxygen-metal bonding states, whereas the second
peak at 526~eV is due to more pure O states as well as some mixing of the
O bands with the occupied (Ni,Co) $t_{2g}$ states. In the XAS, two more
features are visible at 530.5~eV and 532~eV, and match up well with the
theoretical (tetrahedral crystal-field split) V $e_g$ and $t_{2g}$ states
respectively. Experimentally, the splitting between these states is found
to be slightly larger for NVO than CVO.  Finally, at higher energies,
mixing between the metal $4sp$ and oxygen electrons yield the broad band
between 536 and 545~eV.  However, the (Ni,Co) states near $E_{\rm F}$
(predicted by theory at near $-1$ and $+1$~eV) are not observed as distinct
features in the experiment. These results indicate that the occupied (Ni,Co)
states overlap in energy more with the O $2p$ bands, and may account for
the higher intensity of the 526~eV XES feature in NVO compared with CVO.
Correspondingly, the higher intensity of the 530.5~eV XAS feature in CVO,
relative to NVO, may indicate the higher number of available unoccupied
Co states compared with Ni. Somewhat surprisingly, this picture bears more
similarity with that of the LDA+U results, \cite{rai2006,rai2007} in which the
(Ni,Co) states were found to be repelled far from $E_{\rm F}$ into the O $2p$
and V $3d$ states. However, within the LDA+U, the electronic structure of
NVO and CVO are quite dissimilar, incompatible with our spectra. In order
to test what kind of correction to the theoretical band structure would be
required to reproduce the experiment, we have shifted the (Ni,Co)
$d$ bands away from $E_{\rm F}$, mimicking the qualitative trend of the
LDA+U, but of much less magnitude. In this model, occupied Co and Ni $d$
states were rigidly shifted by $-0.4$~eV and $-0.9$~eV respectively for CVO
and NVO and unoccupied states were shifted by $+0.5$ and $+0.7$~eV, chosen to
improve the agreement with experiment. After this shift, the center of mass
of the $d$ states in NVO and CVO end up being quite similar at $-1.9$~eV
($+1.4$~eV) and $-1.7$~eV ($+1.4$~eV) for occupied (unoccupied) states
respectively.  This artificial shift of the bands clearly takes no account
of the evolution in hybridization or $d$ bandwidth that would accompany such
a shift in energy. Nevertheless, the results, shown by the dashed lines in
Fig.~\ref{f:cnvo_pdos}, are able to account for the differences in relative
intensity of the two XES and two XAS features, as well as the increased
bandwidth of CVO, and may hint towards the ultimate fate of the (Ni,Co) states.

This situation is reminiscent of that in NiO, in which the early LDA
calculations yielded bands with 
strong Ni $3d$ character near $E_{\rm F}$, resulting in NiO that was
either metallic,
\cite{mattheiss1972} or had a very narrow insulating gap, \cite{terakura1984b}
depending on whether spin polarization was included or not. Subsequent LDA+U
calculations improved the magnitude of the insulating
gap by pushing the occupied Ni $d$ states deep into the O bands.
\cite{anisimov1991} However,
it was not until dynamic correlations were included in the
form of dynamical mean-field theory (DMFT), \cite{kunes2007}
that good agreement with several
spectroscopic methods was
simultaneously obtained from an {\em ab initio} approach. \cite{kurmaev2008}
In DMFT, the Ni $d$
states of NiO are located at an energy intermediate between the LDA and LDA+U.

\begin{figure*}[t!]
\begin{center}
\includegraphics[width=1.0\linewidth,clip]{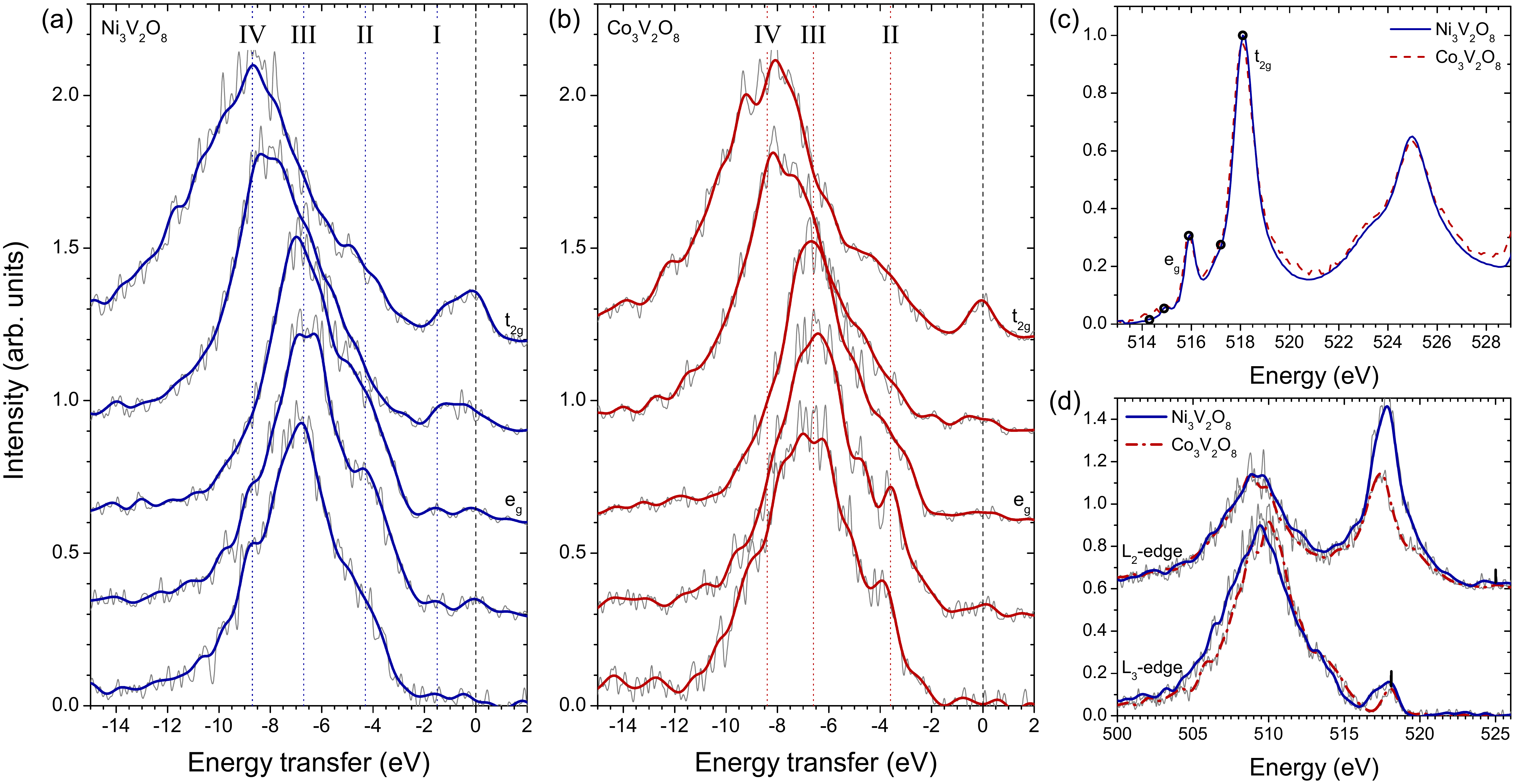}
\end{center}
\vspace*{-0.2in}
\caption{\label{f:cnvo_vl_rixs} (Color online) (a) and (b): V $L_3$-edge RIXS
of NVO
and CVO, vertically offset for clarity. The light gray traces denote the
raw spectra, and the dark solid lines are the noise filtered spectra. The
lowest spectra correspond to the lowest excitation energy.  Roman numerals
indicate the approximate center of the most visible excitations. (c)~XAS
spectra of both compounds, recorded in TEY mode. The open circles denote
the energies used in the RIXS measurements of (a) and (b). (d)~XES spectra
of CVO and NVO at the V $L_2$- and $L_3$-edges.  The vertical bars indicate
the excitation energies of the individual measurements.}
\end{figure*}

\subsection{V $L_3$-edge RIXS}
\label{s:cnvo_vl_rixs}
Figure \ref{f:cnvo_vl_rixs} presents RIXS [Figs.~\ref{f:cnvo_vl_rixs}(a,b)]
XAS [Fig.~\ref{f:cnvo_vl_rixs}(c)] and XES [Fig.~\ref{f:cnvo_vl_rixs}(d)]
spectra recorded near the V $L$ edge.
V $L_{3,2}$-edge XAS spectra are shown in Fig.~\ref{f:cnvo_vl_rixs}(c) for
NVO and CVO. Below $\sim 521$~eV, the double-peaked feature represents
absorption from the V $2p_{3/2}$ core level into unoccupied $e_g$ and $t_{2g}$
states respectively. At higher energies, similar, although broader, features
represent absorption from the $2p_{1/2}$ level. These spectra are very similar
for the two compounds, and are in good agreement with CFM calculations of
tetrahedrally co-ordinated V$^{5+}$.

V $L_3$-edge RIXS spectra of NVO and CVO are shown in
Figs.~\ref{f:cnvo_vl_rixs}(a) and \ref{f:cnvo_vl_rixs}(b) respectively,
and were recorded with the photon and
spectrometer resolution both set to 0.69~eV. Since both compounds are
nominally $d^0$ V$^{5+}$ compounds, XES and RIXS are weak at this edge,
and the recorded spectra suffer from an appreciable noise ratio. In
order to enhance the signal-to-noise ratio, we employ the maximum entropy
method as a noise filter, as discussed above.  Five spectra were recorded
across the V $L_3$-edge of both NVO and CVO at incident energies shown in
Fig.~\ref{f:cnvo_vl_rixs}(c); both the raw spectra and noise filtered spectra
are presented in Fig.~\ref{f:cnvo_vl_rixs}.  The V $L_3$-edge RIXS spectra
have very similar shapes for the two compounds, reflecting the similarity
of the V environment of NVO and CVO, and are dominated by charge-transfer
(CT) excitations between O $2p$ and V $3d$ sites: ${\rm V}\;3d^n\;{\rm
O}\;2p^6 \rightarrow {\rm V}\;3d^{n+1}\;{\rm O}\;2p^5$. For the
first three excitation energies shown in Fig.~\ref{f:cnvo_vl_rixs},
the features are constant on the energy transfer axis, indicating their
origin as loss (CT) features.  The association of these loss features
with CT excitations is consistent with similar observations in other
vanadium oxides.
\cite{duda2004etc}
These spectra correspond to excitations into the unoccupied V $e_g$ states,
and the dominant feature (III) at 6.7~eV for NVO (6.6~eV for CVO) can be
associated with the separation between the center of mass of the O $2p$
states and the unoccupied V $e_g$ band. At higher excitation energies,
corresponding to excitations into the unoccupied V $t_{2g}$ states, the
spectra shift to deeper energies, centered at 8.7~eV (IV) for NVO (8.4~eV
for CVO). The energy separation (of approximately 2~eV) between these two
different CT excitations is consistent with the separation between V $e_g$
and $t_{2g}$ features in the XAS (Fig.~\ref{f:cnvo_pdos}), in particular the
slightly smaller splitting of these features for CVO compared with NVO.

In order to establish the origin of the V $L_3$-edge spectral features,
it is useful to compare them with $L_2$-edge measurements. These spectra
are shown in Fig.~\ref{f:cnvo_vl_rixs}(d) for both CVO and NVO on an emission
energy scale, alongside the resonant $L_3$-edge spectra [the topmost spectra
of Figs.~\ref{f:cnvo_vl_rixs}(a,b)] for comparison. At the $L_2$-edge, emission
in the energy range 507~--~512~eV represents fluorescent emission from
V-O hybridized states into the V $2p_{3/2}$ core level, rather than CT or
$dd$ excitations that are present in the $L_3$-edge spectra at this energy
range. For CVO, there is a clear difference in the shape and center of mass
of this feature between the $L_2$ and $L_3$-edge spectra, supporting our
interpretation that the $L_3$-edge RIXS features (in particular, features
III and IV) are associated with CT inelastic scattering processes. The same
is also true of NVO, albeit less obviously owing to the precise choice of
excitation energy in the NVO $L_3$-edge RIXS spectrum. We also note that
the NVO fluorescent feature is slightly broader than that of CVO, indicating
the bandwidth of the V-O hybridized states of NVO is broader than that of CVO.

\begin{figure*}[t!]
\begin{center}
\includegraphics[width=1.00\linewidth,clip]{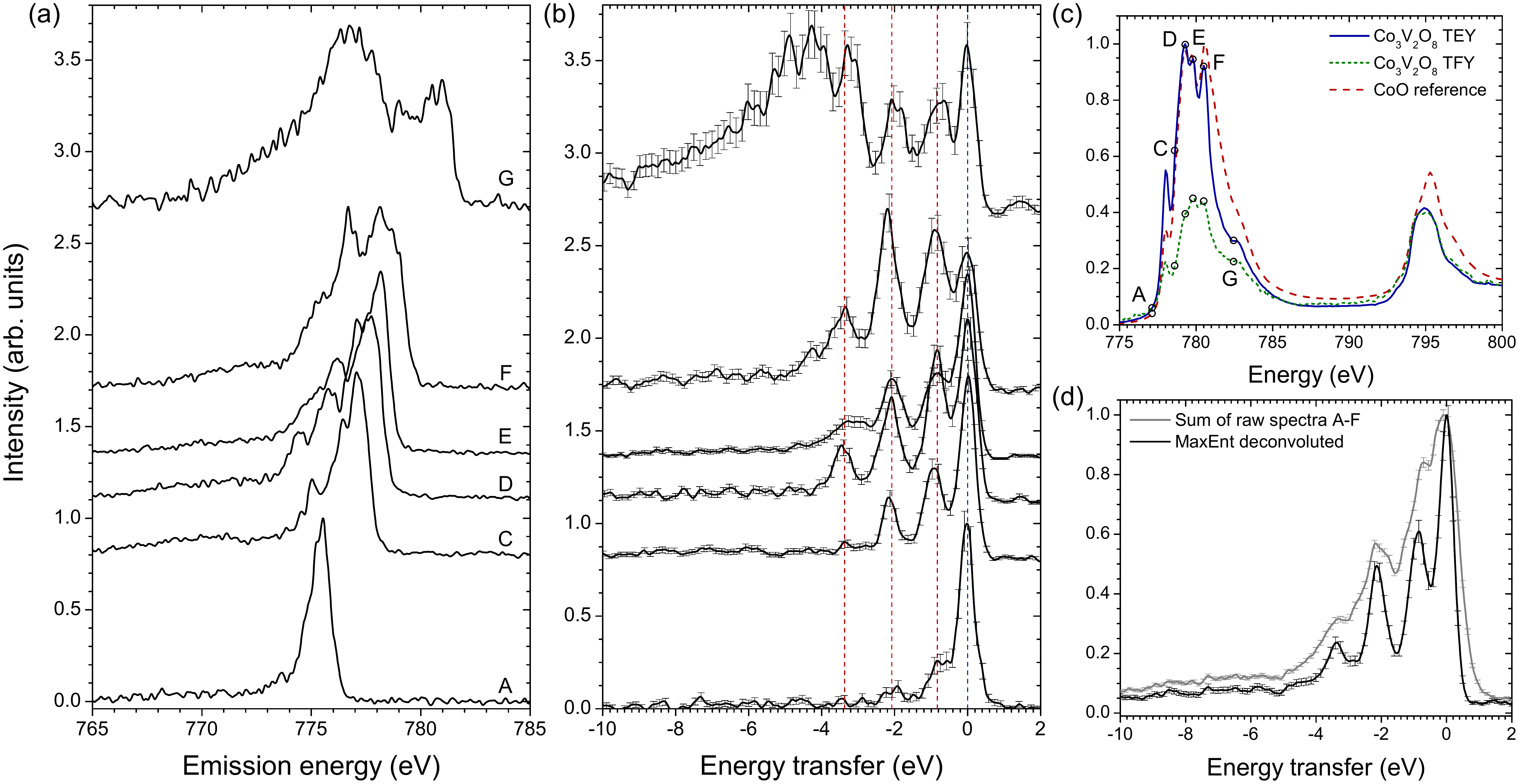}
\end{center}
\vspace*{-0.2in}
\caption{\label{f:cvo_col_rixs} (Color online) Co $L_3$-edge RIXS measurements
of CVO: a) raw spectra on an emission energy scale, and (b) MaxEnt deconvoluted
spectra, shown on an energy transfer scale. (c) Co $L$-edge XAS spectra of CVO,
in which the open circles denote the RIXS excitation energies. (d) Summed RIXS
spectra (A-F), before and after the MaxEnt deconvolution.}
\end{figure*}

Returning to the V $L_3$-edge RIXS, in both compounds an additional low-energy
shoulder (II) is present at lower excitation energies, at 4.3~eV for NVO and
3.6~eV for CVO. Optical measurements have found low-energy CT excitations at
3.0~eV and 4.4~eV for NVO, interpreted as excitations into a
mix of unoccupied Ni/V states and into pure V states respectively.
\cite{rai2006} For CVO,
these optical excitations are found to be shifted to slightly lower energies,
at 2.7~eV and 4.2~eV. \cite{rai2007} Unlike optical measurements, however,
RIXS is a site- and orbital-selective probe of the local transitions, and the
absence of the $\sim 3$~eV feature in our data unambiguously indicates its
origin is not primarily unoccupied V states. We note, however, that this
energy approximately coincides with the onset of the broad CT feature, and
its origin may be weak mixing of V states with (Co,Ni) states, as suggested
by Rai {\em et al.} \cite{rai2006,rai2007} The $\sim 4$~eV feature (II) in our
data, however, is in good agreement with the optical data. Here, we find
it is most intense for low excitation energies, suggesting it represents an
excitation into the empty V $e_g$ band.

Finally, we note that a small peak (I) is apparent in all the NVO spectra
[Fig.~\ref{f:cnvo_vl_rixs}(a)] at an energy of 1.5~eV. This feature is also
present in the V $L_2$-edge XES spectrum of NVO,
visible as a weak peak approximately 1.5~eV offset from the incident photon
energy. The energy of this feature is too low to be a CT transition, and
we instead interpret it as a weak $dd$ transition between partially occupied
and unoccupied V $e_g$ states. The absence of any such peak in the CVO
spectra suggests that the V ion in CVO is more strictly $d^0$, whereas the
extra electron in NVO at least weakly occupies the V $3d$ states.

In summary, our V $L_{3,2}$-edge RIXS measurements identify three CT
transitions present in CVO and NVO. Transitions II and III are CT excitations
into empty V $e_g$ states, whereas IV involves empty V $t_{2g}$ states.
Transitions III and IV occur at similar energies, of $\sim 6.6$~eV and $\sim
8.5$~eV respectively, for both CVO and NVO. Transition II, on the other hand,
is at 3.6~eV for CVO and 4.3~eV for NVO, and may involve V $e_g$ states
that are mixed with unoccupied (Co,Ni) $3d$ states, in rough agreement with
the optical measurements. \cite{rai2006,rai2007} In addition to these CT
transitions, a $dd$ excitation is also observed for NVO at 1.5~eV for all
incident energies across both $L_3$ and $L_2$ edges, and is interpreted as
being from occupied to unoccupied $e_g$ states. The absence of
this excitation for CVO suggests the V $3d$ states are negligibly occupied
in CVO.

\begin{figure}[t!]
\begin{center}
\includegraphics[width=0.90\linewidth,clip]{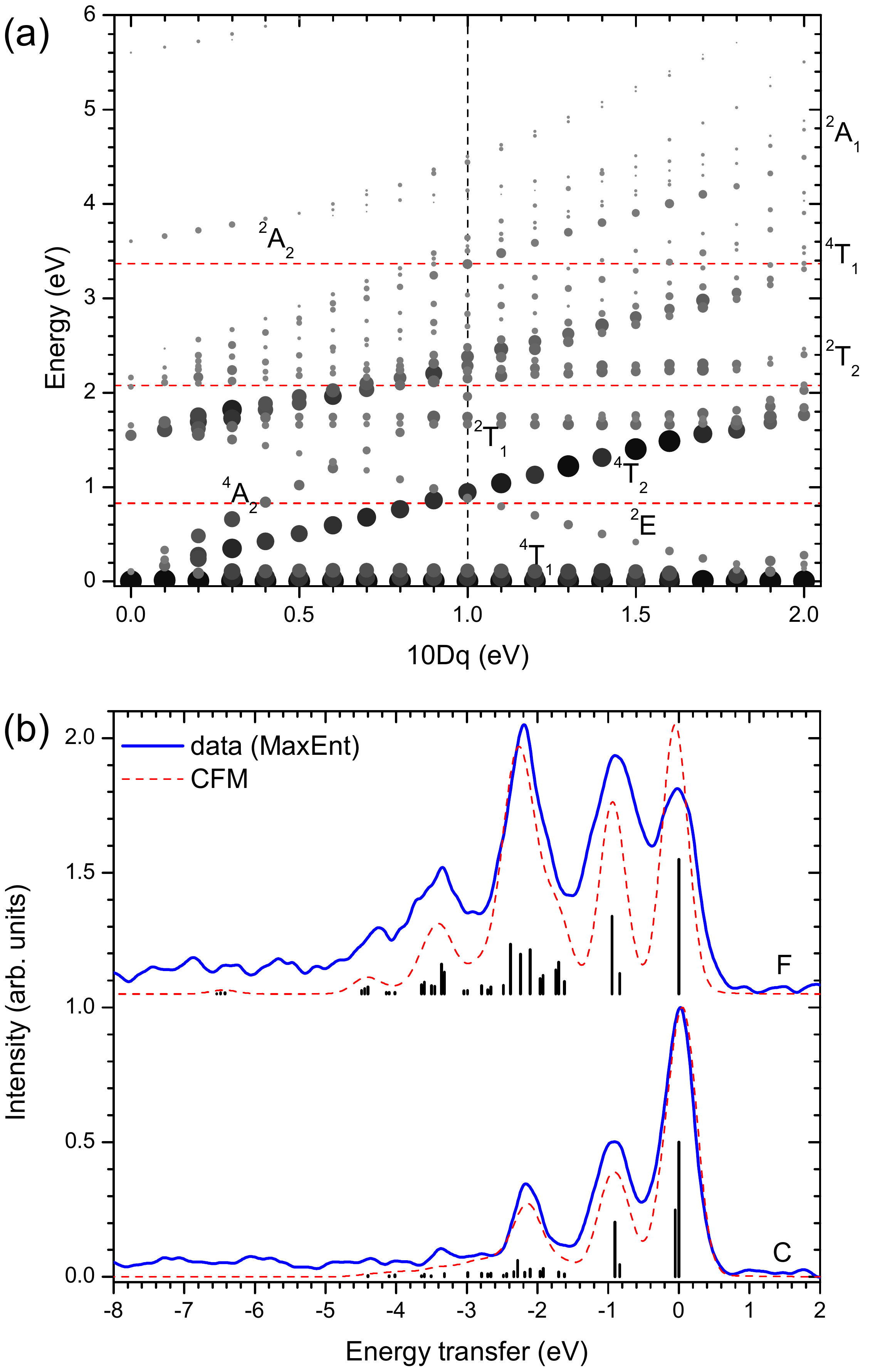}
\end{center}
\vspace*{-0.2in}
\caption{\label{f:cots} (Color online) CFM calculations of the Co$^{2+}$
$d^7$ ion. (a)~Tanabe-Sugano--like diagram of Co$^{2+}$ in $O_h$ symmetry
with a 65\% reduction of the Slater integrals. The relative intensity of
the features in $L_3$-edge RIXS measurements, predicted by the CFM model,
is depicted by the size and shade of the points. Horizontal dashed lines
indicate the experimental features identified in Fig.~\ref{f:cvo_col_rixs}(b).
(b)~Comparison of CFM calculations ($10Dq = 1.0$~eV) with the experimental
data of spectra C and F. The discrete transitions
are shown by the sticks, whereas the dashed line has been convoluted with
the experimental resolution function.}
\end{figure}

\subsection{Co $L_3$-edge RIXS}
\label{s:cvo_col_rixs}
RIXS measurements at the Co $L_3$-edge of CVO are shown in
Figs.~\ref{f:cvo_col_rixs}(a,b), at the excitation energies shown in the XAS
spectrum
of Fig.~\ref{f:cvo_col_rixs}(c). Six spectra were recorded in the second order
of diffraction with a photon and spectrometer resolution of 0.4~eV and
0.82~eV respectively, and are presented in Fig.~\ref{f:cvo_col_rixs}(a).
Owing to the higher count rate of the RIXS processes at
the Co $L$-edge compared with the V $L$-edge in CVO, the MaxEnt procedure
is used as a deconvolution tool (rather than as a noise filter).
The spectra were deconvoluted using a broadening function of
0.82~eV FWHM, equivalent to the spectrometer resolution function, and are
shown in Fig.~\ref{f:cvo_col_rixs}(b). The errorbars
in Fig.~\ref{f:cvo_col_rixs}(b) are obtained through the empirical relation,
$\sigma_i = N_i^{(1/1.54)}$ (where $N_i$ is the intensity of the datum $i$
of the deconvoluted spectrum), determined from a detailed investigation into
the propagation of noise through the MaxEnt procedure. \cite{laverock2011b}

Aside from the rather intense elastic peak at 0~eV, the spectra are
dominated by scattering at energy transfer of less than 5~eV, which can
be attributed to $dd$ excitations of the Co $d$ electrons. Above 5~eV,
a weak band is present, centered at $6.5$~eV, which is due to Co-ligand
CT transitions. Three $dd$ features are clearly identifiable, even in the
raw spectra [Fig.~\ref{f:cvo_col_rixs}(a)], at (i)~0.9~eV, (ii)~2.1~eV and
(iii)~3.4~eV. In the deconvoluted spectra [Figs.~\ref{f:cvo_col_rixs}(b,d)],
these features become well separated and are easily visible. A fourth feature
at $\sim 4.6$~eV (iv) may be present for some of the higher excitation energies
(e.g.~spectra F-G). These spectra are reminiscent of RIXS
measurements of other octahedrally-coordinated Co$^{2+}$ compounds. For
example, in high-resolution Co $L$- and $M$-edge measurements of CoO,
Chiuzb\v{a}ian {\em et al.}\ found a total of five transitions at 0.9,
1.9, 2.3, 3.0 and 3.5~eV, although the 1.9 and 2.3~eV features were not
resolvable from one another in their $L$-edge data. \cite{chiuzbaian2008}

In order to understand the Co $L_3$ RIXS spectra more thoroughly, CFM
calculations have been performed for the Co$^{2+}$ ion, paying particular
attention to the intensity dependence of the transitions with excitation
energy.  The dependence of the crystal-field excitations in $O_h$ symmetry as a
function of the crystal-field splitting, $10Dq$ is shown in Fig.~\ref{f:cots}(a)
for a reduction of the Slater integrals to 65\% of the Hartree-Fock (HF)
values. In many $3d$ systems, a reduction of the Slater integrals to 80\%
of the HF values is required to accurately describe multiplet effects in
XAS and RIXS. \cite{degroot2005} Here, we find that
this more modest reduction of the Slater parameters leads to $dd$ transition
energies that are $\sim 10$\% too large. Neither adjusting $10Dq$ or lowering
the symmetry to $D_{4h}$ were able to significantly improve the agreement.
Although larger reductions (to 75\%) have been obtained through careful
fitting of RIXS data to CFM results, \cite{ghiringhelli2006} we do not
consider the current reduction as representative of this system. Rather,
it is more likely that it serves to adequately counteract other more severe
limitations of our model calculations. In particular, calculations in the
correct (lower) symmetry may more accurately reflect the data. However,
without the rich spectral structure afforded by high-resolution measurements,
it is difficult to optimize the fine splittings that are present at lower
symmetry; we therefore restrict our analysis to $O_h$ symmetry here, which
at least captures the qualitative behavior well.

\begin{figure*}[t!]
\begin{center}
\includegraphics[width=1.00\linewidth,clip]{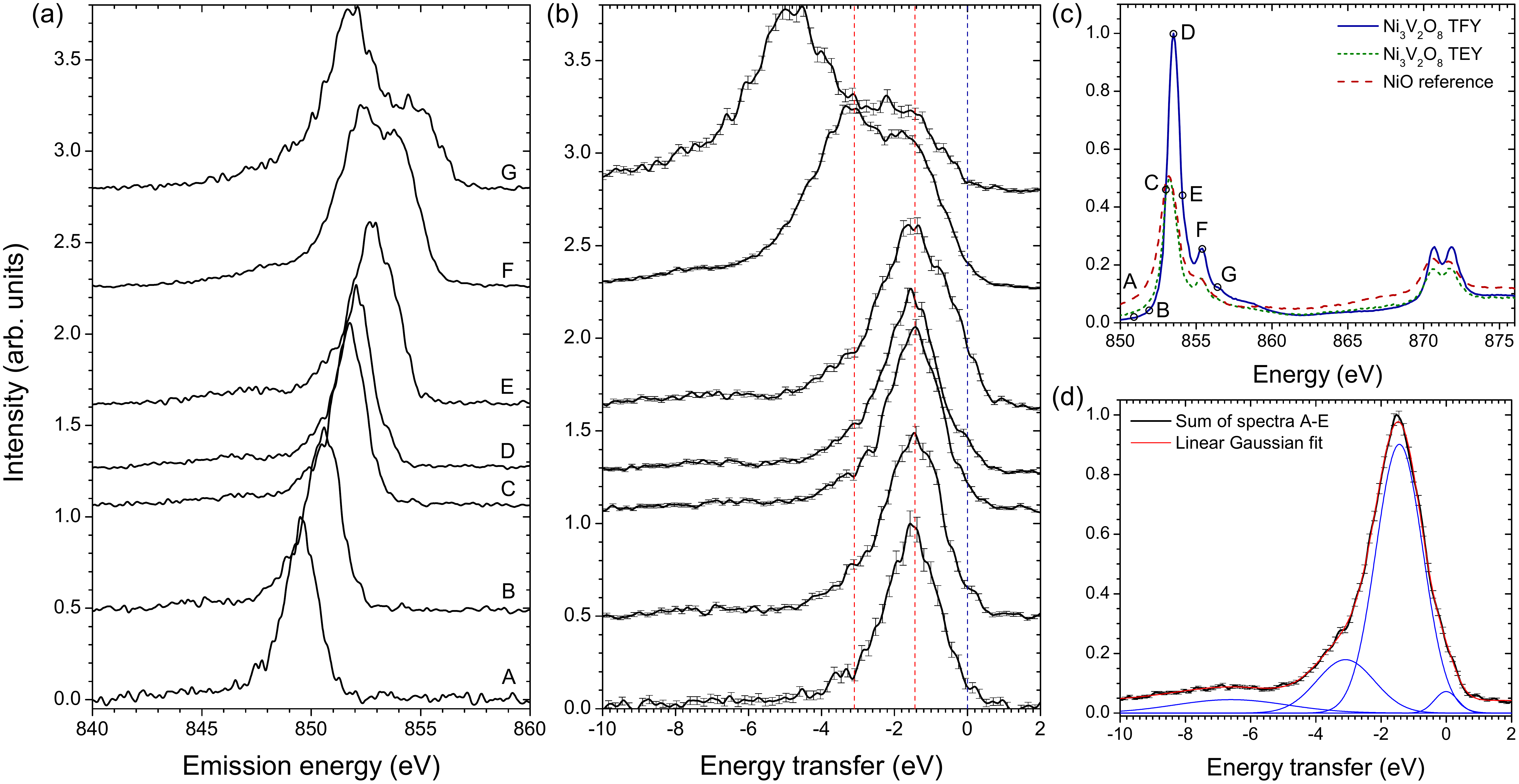}
\end{center}
\vspace*{-0.2in}
\caption{\label{f:nvo_nil_rixs} (Color online) Ni $L_3$-edge RIXS measurements
of NVO: raw spectra shown on (a)~an emission energy and (b)~a loss energy
scale. (c)~Ni $L$-edge XAS spectra showing the excitation energies
chosen for the RIXS measurements as open circles. (d) Summed RIXS spectra
A-E, and the results of fitting the summed spectra to a linear combination
of four Gaussian functions.}
\end{figure*}

The energies and relative intensities in Co $L_3$-edge RIXS of the respective
$dd$ transitions are shown in Fig.~\ref{f:cots}(a); the intensities are shown
as the average intensity across the Co $L_3$-edge. Spin-orbit coupling,
known to be important in the CoO multiplet structure, is included, and
splits the $^4T_1$ ground state into four sub-states: the symmetry labels
refer to the symmetry of the states without spin-orbit coupling. For $10Dq
\approx 1.0$~eV, the calculated $dd$ transitions approximately intersect
our experimental features. Moreover, these transitions have the correct
intensity dependence with the excitation energy. For example, the $^4T_2$
state is the only transition predicted by the CFM calculation to weakly
resonate at the onset of the XAS spectrum A. At energy C, the $^4T_1$
configuration contributes to the spectrum, providing intensity near 2~eV in
our measurements. Explicit comparison between our data and the CFM model
is shown in Fig.~\ref{f:cots}(b) at two representative excitation energies,
demonstrating the good agreement. In these calculations, the sub-states
of $^4T_1$ are populated by a thermal (Boltzmann) distribution at room
temperature.

With the aid of the CFM model, we assign (i) to excitations to the $^4T_2$
state, (ii) to a combination of $^4T_1$ and $^4A_2$ (with some weak intensity
from the $^2T_1$ and $^2T_2$), (iii) to the spin-orbit split $^2A_1$
configuration, and (iv) to the weak $^2A_2$ transition. Our assignment
(which improves on an earlier preliminary one \cite{laverock2011b})
is similar to that of CoO, \cite{magnuson2002} emphasizing the
local nature of the RIXS processes at the Co $L$-edge in CVO.

\subsection{Ni $L_3$-edge RIXS}
\label{s:nvo_nil_rixs}
RIXS measurements across the Ni $L_3$-edge of NVO are shown in
Figs.~\ref{f:nvo_nil_rixs}(a,b) at the excitation energies shown in the
absorption
spectrum of Fig.~\ref{f:nvo_nil_rixs}(c).  Although Ni $L_{3,2}$ XAS has been
recorded, it is presented here only as a guide to the RIXS measurements, owing
to the presence of elemental Ni in the upstream x-ray optics of ALS BL7.
Absorption
of the incident x-rays at the Ni $L$-edge means knowledge of the incident
photon flux at the sample is not reliable. The precise shape of the spectrum
is therefore difficult to quantify, and no attempt is made here to interpret
these results, other than to emphasize they agree with a +2 charge state
for Ni in NVO. We note, however, that although XAS measurements at the Ni
$L$-edge are somewhat hampered, the effect on RIXS is only to renormalize
the incident photon flux, and there will be no effect on the shape of the
individual RIXS spectra.

\begin{figure}[t!]
\begin{center}
\includegraphics[width=0.90\linewidth,clip]{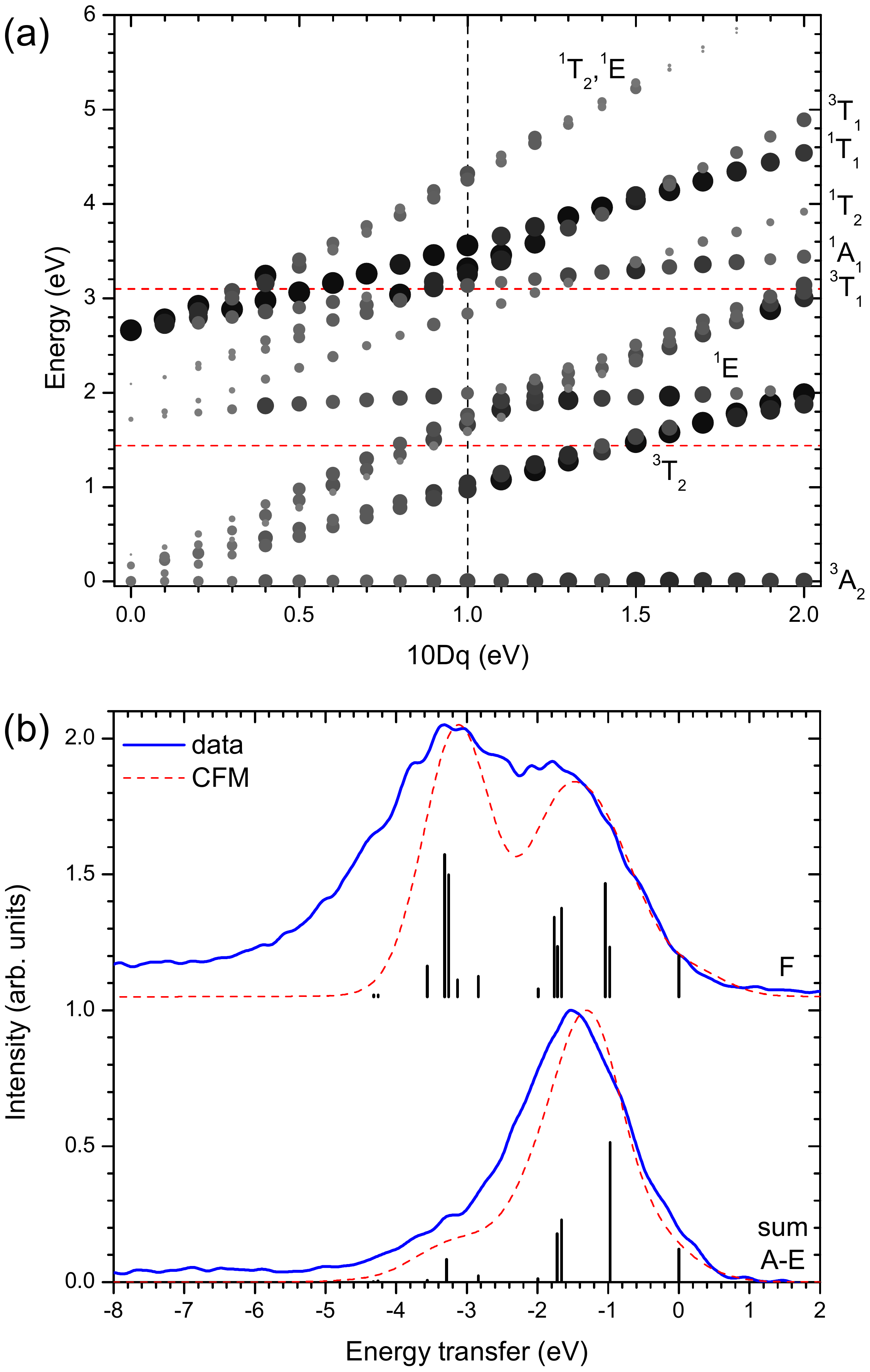}
\end{center}
\vspace*{-0.2in}
\caption{\label{f:nits} (Color online) CFM calculations of the Ni$^{2+}$ $d^8$
ion. (a)~Tanabe-Sugano--like diagram of Ni$^{2+}$ in $O_h$ symmetry. The
relative intensity of the features in $L_3$-edge RIXS measurements,
predicted by the CFM model, is depicted by the size and shade of the
points. Horizontal dashed lines indicate the experimental features identified
in Fig.~\ref{f:nvo_nil_rixs}(d).  (b)~Comparison of CFM calculations ($10Dq =
1.0$~eV) with the experimental data of spectrum F and the sum of spectra
A-E. The discrete transitions are shown by the sticks, whereas the dashed
line has been convoluted with the experimental resolution function.}
\end{figure}

Seven RIXS spectra were recorded in the second order of diffraction [shown
in Figs.~\ref{f:nvo_nil_rixs}(a,b)] at photon and spectrometer resolutions
of 0.50 and 0.97~eV respectively for spectra A-E, and at 0.8 and 1.7~eV
for spectra F-G. Unlike CVO, elastically scattered light is substantially
suppressed for NVO in this scattering geometry, and is barely visible in
the raw data, only becoming resolvable after deconvolution (not presented
here).  This strong suppression of the elastic peak is in agreement with
$L$-edge RIXS measurements of NiO, \cite{ghiringhelli2009betc}
in which the Ni$^{2+}$ ion is also octahedrally co-ordinated. Indeed, the
structure of our measurements are reminiscent of both these $L$-edge
measurements, \cite{ghiringhelli2009betc} as well as $M$-edge
measurements of NiO.\cite{chiuzbaian2005} The most prominent feature
of the data is the strong dispersing peak in Fig.~\ref{f:nvo_nil_rixs}(a),
which is found to be centered at an energy of 1.44~eV on the loss energy
scale of Fig.~\ref{f:nvo_nil_rixs}(b). In Fig.~\ref{f:nvo_nil_rixs}(d), the
sum of spectra A-E (which are unambiguously dominated by RIXS processes)
is shown, alongside a fit to a linear combination of Gaussian functions
(a total of four Gaussians were found to be required to adequately describe
the data). This analysis clearly reveals the presence of the weak elastic
peak, as well as the weak shoulder on the low-energy side of the main peak
at 3.10~eV, which becomes strong when the excitation energy is tuned to
the main satellite in the absorption spectrum (energy F). In addition,
a broad band of excitations is observed centered at 6.6~eV, representing
Ni-ligand charge-transfer transitions. This fit also provides information on
the widths of the RIXS features. While the fitted elastic peak has a width
(0.85~eV FWHM) comparable with that expected from the experimental setup
($\sim 1.1$~eV), the widths of the two low-energy transitions are much broader
($\sim 2$~eV). Indeed, the widths of the RIXS features in NVO are visibly
larger than in CVO, despite the similar total experimental resolution in
both measurements ($\sim 0.9$~eV for CVO), an effect we return to in more
detail below. Although the MaxEnt deconvolution was attempted on these data,
the large intensity difference between spectral features was found to limit
the power of the deconvolution, with the result that the deconvoluted spectra
were only found to sharpen, without separating individual features.

In order to understand the Ni $L_3$ RIXS spectra more thoroughly, CFM
calculations have been performed in $O_h$ symmetry, with a reduction of the
Slater integrals to 80\% of their atomic values. Owing to the resolution of
these measurements, the optimization of the crystal-field parameters has
not been attempted. In particular, the distortion of the local environment
of the Ni ion lowers the symmetry, and therefore breaks the degeneracy of
(and splits) the transitions.  In practice, however, such a distortion has
a weaker impact on the transition spectrum, compared with the resolution
of the current measurements.  In Fig.~\ref{f:nits}(a), the dependence of
the CFM calculation on the crystal-field parameter, $10Dq$, is presented
as a Tanabe-Sugano--like diagram.  The relative intensity of the features
predicted by the CFM model has been averaged from just above the $L_3$
threshold energy (approximately between points D and E) to point G in
Fig.~\ref{f:nvo_nil_rixs}(c), owing to the overestimation of the intensity of
the elastic feature for lower excitation energies in the CFM calculation. We
note that SIAM calculations, explicitly including the effects of the ligand,
account well for the weak elastic feature across the entire $L_3$ edge of
NiO. \cite{ghiringhelli2009betc}  For $10Dq \approx 1.0$~eV, a group
of transitions intersect our two experimental features. In Fig.~\ref{f:nits}(b),
the experimental spectra are compared directly with the CFM calculation, after
broadening with the experimental resolution function. Within the resolution
of the measurement, reasonable agreement is observed; in particular, this
comparison demonstrates that our 1.44~eV feature is composed of several
unresolved transitions. Indeed, it is possible to see some asymmetry of this
feature in the raw spectra. For example, spectrum E has a rather prominent
shoulder centered at $\sim 0.9$~eV.

Overall, with the aid of the CFM model, we associate the 1.44~eV feature with
a combination of $^3T_2$ and $^3T_1$ transitions (the $^1E$ transition is
substantially weaker for most excitation energies). Similarly, the 3.10~eV
feature is dominated by $^3T_1$ and $^1T_1$ transitions (with some weak
intensity from transitions of $^1T_2$ and $^1A_1$ symmetry). Finally, we
note that some evidence of the higher energy $^1T_2$, $^1E$ excitations may
be visible in spectrum E at an energy transfer of $\sim 4.1$~eV.

\section{Discussion}
Summarizing the experimental results, we find that the electronic structure of
{\em M}$_3$V$_2$O$_8$ ({\em M}~=~Ni, Co) differs from the predictions of the
LSDA primarily in the location of the (Ni,Co) $3d$ states, which are placed
too close to $E_{\rm F}$. Rigidly moving apart the occupied and unoccupied $d$
bands by 1 -- 2~eV captures some of the features of the data, and indicates
the energetic location of the center of mass of these states: occupied
states at $\sim -1.8$~eV are located near the high-energy feature of the XES
spectrum, and unoccupied states at $\sim +1.4$~eV are close to the V $e_g$
states of the XAS. This trend (to push apart the occupied and unoccupied
$d$ bands) is similar to the results of the LSDA+U method, although the
magnitudes of the shifts employed here are much smaller. Indeed, previously
published LSDA+U results yield quite different electronic structures for NVO
and CVO, which is not experimentally observed.  Presumably, more sophisticated
calculations, such as DMFT, may be capable of more faithfully reproducing the
experimental spectra, and in particular the (Ni,Co) $d$ electron energies.
The V and O states, on the other hand, are relatively insensitive to the
choice of {\em M} and agree reasonably well with experiment. However, there is
some evidence of a V $dd$ transition at 1.5~eV in NVO, which is not observed in
CVO, and which may indicate some weak occupation of the V $3d$ states in NVO.

Additional circumstantial evidence for this picture comes from the (Ni,Co)
$L_3$ RIXS measurements. Charge transfer excitations between occupied O $2p$
states and unoccupied (Ni,Co) $d$ states are observed centered at 6.6~eV and
6.5~eV for NVO and CVO, respectively. These are very similar in magnitude to
the V $e_g$ CT excitations identified in V $L_3$ RIXS (6.6~eV and 6.7~eV),
suggesting the unoccupied (Ni,Co) $d$ states and V $e_g$ states lie close
in energy, as indicated by their centers of mass. In both cases, it is
the bonding O $2p$ peak (lower peak in the XES spectrum) that is $\sim
6.5$~eV below. Additionally, we note that V $t_{2g}$ CT excitations are
observed $\sim 2$~eV above the $e_g$ CT excitations (and in agreement with
the XAS data), supporting our assignment. Finally, a third CT excitation is
evident in the V $L_3$ RIXS at 4.3~eV (3.6~eV), and is interpreted as from
the weakly-bonded O $2p$ states into empty V $e_g$ states, in agreement with
optical measurements of CT features. \cite{rai2006,rai2007}

At the (Ni,Co) $L$-edges, RIXS measurements reveal $dd$ transitions in good
agreement with crystal-field multiplet calculations with a crystal-field
splitting of $10Dq \approx 1$~eV, and are strongly reminiscent of the
corresponding binary oxides, NiO and CoO. The experimental features in
NVO are associated with crystal-field transitions to the $^3T_2$/$^3T_1$
(1.44~eV), $^3T_1$/$^1T_1$ (3.1~eV) and $^1T_2$/$^1E$ (4.1~eV) symmetry states
respectively.  For CVO, the experimental $dd$ transitions are assigned to
the $^4T_2$ (0.9~eV), $^4T_1$/$^4A_2$ (2.1~eV), $^2A_1$ (3.4~eV) and $^2A_2$
(4.6~eV) excited states.

As mentioned above, there are two distinct (Ni,Co) sites within the unit
cell, referred to as cross-tie (Ni1,Co1) and spine (Ni2,Co2). It has
been suggested from optical measurements that the
crystal-field environment may be quite different between the two sites,
even in the paramagnetic phase at room temperature, leading to optical $dd$
transitions separated by 0.6~eV (0.75 and 1.35~eV) and 0.9~eV (0.7 and 1.6~eV)
for NVO and CVO respectively, with the separation being more severe for the Co
compound. \cite{rai2006,rai2007}
In our band structure calculations, we do not find strong differences
in the crystal-field splitting between spine and cross-tie sites. Indeed,
the center of mass of the $t_{2g}$ and $e_g$ states at the two different
sites agree to within 0.04~eV for both compounds (i.e.~more than an order
of magnitude smaller than suggested by optical measurements). Although the
accuracy of the LSDA must be questioned for these compounds, this optical
splitting is also not observed in our experimental RIXS spectra, which are
found to agree very well with atomic multiplet theory of a single value of
the crystal-field splitting.  Whilst we cannot exclude such splittings of
$\lesssim 0.2$~eV, splittings $>0.5$~eV ought to be directly visible in our
spectra. For both compounds, the lowest optical transitions ($\sim 0.7$~eV)
are close to those identified above via RIXS. However, for CVO the second
optical feature (at 1.6~eV) lies in a dip in our RIXS spectra.  RIXS is a
direct probe of on-site electronic transitions, whereas their detection in
optical measurements relies on hybridization with neighboring ligands. The
absence of the optical features in RIXS measurements may indicate that
their origin is more complex, possibly involving dimer excitations. Dimer
excitations, involving two neighboring Co sites and of the form $\{d^7;\;d^7\}
\rightarrow \{d^8;\;d^6\}$, are allowed in optical measurements, but are
usually expected to be weaker than single-site excitations in RIXS. Their
energy can be approximately estimated through the relation, $U_{\rm eff} -
J$ which, for reasonable values of $U_{\rm eff}$ and $J$, would make them
relevant at 1~--~2~eV, approximately the right energy regime.

\section{Conclusion}
In summary, we have measured the electronic structure of the local
moment insulators {\em M}$_3$V$_2$O$_8$ ({\em M}~=~Ni, Co) using soft x-ray
absorption, emission and resonant inelastic scattering. Whereas the V and O
states are in reasonable agreement with {\em ab initio} FLAPW band structure
calculations, the LSDA fails to account well for the Ni and Co sites. The
location of the narrow Ni and Co $3d$ bands is predicted to lie close to
(and, in the case of Co, cross) $E_{\rm F}$, which is not experimentally
observed. On the other hand, features in the O $K$-edge XES and XAS indicate
the energetic location of these bands, which appear deeper into the occupied O
$2p$ and unoccupied V $3d$ manifolds than theoretically expected. By
shifting apart the occupied and unoccupied states, we estimate the centers
of mass of the (Ni,Co) $d$ states.  The weak occupation of the V $e_g$
states in NVO, absent for CVO, may hint towards the different effects of
electron localization in the two materials. In that respect, and considering
the unsatisfactory agreement between both LDA and LDA+U and experiment
reported elsewhere both above and below $T_{\rm N}$, \cite{rai2006,rai2007}
{\em M}$_3$V$_2$O$_8$ would likely benefit from more thorough treatment of
electron correlations, such as that afforded by DMFT (although it is
acknowledged that such a treatment for these materials is not trivial).
On the other hand,
we find good agreement between (Ni,Co) $L$ RIXS and crystal-field multiplet
theory, suggesting the crystal field is not as strongly distorted between
the cross-tie and spine sites as has been previously suggested. Together,
these measurements put strong constraints on the nature of any theoretical
treatment of {\em M}$_3$V$_2$O$_8$.

\section*{Acknowledgements}
The Boston University program is supported in part by the Department of Energy
under Grant No.\ DE-FG02-98ER45680. The ALS, Berkeley, is supported by the
U.S.\ Department of Energy under Contract No.\ DE-AC02-05CH11231. The NSLS,
Brookhaven, is supported by the U.S.\ Department of Energy under Contract
No.\ DE-AC02-98CH10886. The research programme at the University of Warwick is
supported by EPSRC, UK (EP/I007210/1).

\end{document}